\documentclass[twocolumn,showpacs,amsmath,amssymb]{revtex4}
\input epsf
\usepackage{rotating}
\begin{document}
\title {Two dimensional dipolar scattering with a tilt}
\author{Christopher Ticknor}
\affiliation{Theoretical Division, Los Alamos National Laboratory, Los Alamos, New Mexico 87545, USA}
\date{\today}
\begin{abstract}
We study two body dipolar scattering in two dimensions with a tilted
polarization axis.  This tilt reintroduces the anisotropic interaction in a 
controllable manner. As a function of this polarization angle we present 
the scattering results in both the threshold and semi-classical regimes. 
We find a series of resonances as a function of the angle 
which allows the scattering to be tuned.  However the character of the 
resonances varies strongly as a function angle.
Additionally we study the properties of the molecular bound states as a
function of the polarization angle.  
\end{abstract} 

\pacs{34.20.Cf,34.50.-s,34.20.-b}
\maketitle
\section{introduction}
There are exciting proposals based on dipolar gases in two-dimensional
(2D) geometries, such theories show dipolar systems will lead to exotic 
and highly correlated quantum systems \cite{bar}. There has also
been tremendous advances in the production of ultracold
polar molecules \cite{carr}, especially at JILA where a group has 
produced ultracold RbK in 1D optical lattice \cite{fermi-q2d}.
The group used an electric field to align the molecular dipole moments
($\vec d$) along the $\hat z$ direction perpendicular to the plane of 
motion ($\vec \rho$). The tight trapping geometry and 
the dipolar interaction were used to 
inhibit the molecules from reaching their short range interaction where 
they would chemically react \cite{silke}.  

There are alternative molecular systems which will not chemically react 
and less restrictive configurations can be considered. 
For example RbCs and NaCs are chemically stable \cite{chemstab}.  
One interesting possibility to control the properties
of the these dipolar gases is to tilt the 
polarization axis into the plane of motion. 
Such a scenario has been considered by many-body theories: 
for example anisotropic superfluidity has been predicted  \cite{aniso}, 
2D dipolar fermions have been studied \cite{tilt-fermi}, and few body 
dipolar complexes have been investigated \cite{wigner-few}.
However little is known about the nature of the scattering physics of such a
2D system with some in-plane polarization. 
That is the aim of this work; we study the scattering properties of the 
2D dipolar system when the polarization is not fixed out of the plane of motion.  
This reintroduces the anisotropy of the 
interaction {\it controllably}, such that there is a 
preferred direction where the dipoles can line up 
in a head to tail fashion that is energetically favored to the side by 
side configuration, just as in 3D dipolar physics. But in this case the 
strength  of the anisotropy can be controlled directly by the polarization
angle. For small angles, there is only a weaker repulsion in one direction, 
but for large angles (near $\pi/2$) there is an attractive interaction. 

Some recent work has aimed at understanding the scattering 
behavior of dipoles in 2D and quasi-two dimensions (q2D).  First, dipolar scattering
2D in the threshold and semi-classical regime  was studied in pure 2D 
determining the limiting behavior of such scattering \cite{ticknor2d}. 
Then q2D was studied \cite{ticknorq2d} and more recently Ref. 
\cite{jose} introduced an elegant method to solve the full scattering
problem.  Other theories have focused on understanding 
scattering and chemical reactions in q2D \cite{Goulven,micheli}
and how to use the electric field and trap to control the scattering rate.
There has also been some recent work on understanding the scattering
and bound state structure of 2D layer dipolar systems \cite{layer,layer2}
and in a layered system with a tilted polarization axis \cite{volo}.

In the next section we look at the basic scattering of the system 
and offer estimates of the scattering as a function of the polarization 
angle.  Then we look at the character of the scattering resonances.  Finally
we study the molecules that can be formed, their binding energies, size, 
and shape as a function of polarization angle. 

\section{Equations of motion}
For this work we assume that the length scale of confinement is much 
smaller than any other length.  This effectively removes it from the 
scattering problem.  Realistically this length scale will be important, 
but as a first study to provide useful estimates of the scattering this 
assumption is justified.  The Schr\"{o}dinger equation for two 
dipoles in 2D is
\begin{eqnarray}
&&\left(-{\hbar^2\over2\mu}\nabla^2_{\vec\rho}
+d^2{1-3(\hat d \cdot \hat \rho)^2\over \rho^3}
\right)\psi={E}\psi. \label{fullTISE}
\end{eqnarray}
where the dipoles are polarized along 
$\hat d= \hat z \cos(\alpha)+\hat x\sin(\alpha)$ with magnitude $d$ and 
$\vec \rho=x \hat x+y\hat y$.
We solve this equation by using partial wave expansion:
$\psi(\vec\rho)=\sum_m \phi_m(k\rho) e^{im\phi}/\sqrt{\rho}$.
In this case the tilted polarization axis ruins the cylindrical symmetry, 
meaning that different azimuthal symmetries are coupled together.  
Important features of the interaction anisotropy can be distilled by 
looking at the matrix elements:
\begin{eqnarray}
&&\langle m|1-3(\hat d \cdot \hat \rho)^2|m^\prime\rangle=U_{mm^\prime} \nonumber\\
&&=\left(1-{3\over2}\sin^2(\alpha)\right)\delta_{mm^\prime}
-{3\over4}\sin^2(\alpha)\delta_{mm^\prime\pm2}
\label{pot}
\end{eqnarray}
For  $\alpha=0$, the system is totally repulsive and isotropic. Then as 
$\alpha$ is increased the isotropic repulsive term is weakened in the $x$ 
direction but it is still full strength in the $y$ direction.  
This anisotropy enters as couplings between channels with $m$ and
$m\pm2$. At the $\alpha_c=\sin^{-1}(\sqrt{2/3})\sim54.7$ degrees
or $\alpha/\pi\sim0.3$ there is no barrier to the short range and 
past this angle there is an attractive dipolar diagonal potential.

Using the matrix elements in Eq. (\ref{pot}) and the dipolar length 
$D=\mu d^2/\hbar^2$ to rescale Eq. (\ref{fullTISE}), we obtain a 
multi-channel radial Schr\"{o}dinger equation describing 2D dipolar 
scattering with tilted polarization axis:  
\begin{eqnarray}
&&\left(-{1\over2}{d^2\over d\tilde\rho^2}+{m^2-1/4\over 2\tilde\rho^2}
-{E\over E_D}\right)\phi_{m}(\tilde\rho)=\nonumber\\
&&-\sum_{m^\prime}{U_{mm^\prime}\over\tilde\rho^3}\phi_{m^\prime}(\tilde\rho).
\label{TISE}
\end{eqnarray}
where $\tilde\rho=\rho/D$ and $E_D$ the dipolar energy is $\hbar^6/\mu^3d^4$.
To perform the scattering calculation we start at $\rho_0/D$. We vary this 
parameter and use it to control the scattering properties of the system;
$\rho_0/D$ is used to tune the $m$=0 scattering or the scattering length 
$a/D$.  This distance signifies where the interaction 
becomes more complicated through transverse modes  
or system specific interactions becoming important.
Thus a more sophisticated boundary condition is required at this wall. 
However for our initial treatment, we only demand the wave function be
zero at $\rho_0/D$.  

It is worth while to comment that using $\rho_0/D$ to parameterize the 
scattering should be viewed as varying the electric field.
$D$ is $\mu d^2/\hbar^2$ and as and electric field is increased the
induced dipole moment, $d$, becomes larger.  Thus 
decreasing $rho_0/D$ mimics an increasing electric field.  
Additionally, the correspondence of  $rho_0/D$ to $a/D$ 
is unique but due to the nature of the problem (multi-channel) it 
is complex and numerically found. 

Before we present the full numerical scattering calculations,
we discuss the form of the free ($U_{mm^\prime}=0$) 2D wavefunctions
for both scattering and bound states.  The scattering wavefunction is
$\phi_m(k\rho)=\cos(\delta_m)f_m(k\rho)-\sin(\delta_m)g_m(k\rho)$ where 
$\delta_m$ is the  scattering phase shift for the $m$ partial wave and 
$f_m$ $(g_m)$ is the regular (irregular) free solution.  In 2D, it is 
$\sqrt{k\rho}J_m(k\rho)$  ($\sqrt{k\rho}N_m$) where $J_m$ ($N_m$) is a
Bessel (von Neumann) function and $k=\sqrt{2\mu E}$.
If the system were bound then the asymptotic wavefunction for the $m=0$ is 
$\phi_b=\sqrt{\kappa\rho}K_0(\kappa\rho)$  where $K_0$ is the modified 
Bessel function and in the large $\rho$ limit this decays as 
$e^{-\kappa \rho}$ with $\kappa=\sqrt{-2\mu E_b}$ and $E_b$ is the binding 
energy.

As in 3D, the scattering length is defined by when the zero energy 
wavefunction is zero, 
$\psi(a)=\phi_0(a)=0$: $\phi_0(a)=\cot(\delta_0)f_0(a)-g_0(a)=0$ \cite{verhaar}.
Then the scattering length can be computed with
$a={2\over k}e^{{\pi\over2}\cot(\delta_0)-\gamma}$ where $\gamma$ is the Euler 
gamma function $\sim$0.577.
Conversely, the phase shift can be defined by the scattering length: 
$\cot(\delta_0)={2\over\pi}(\ln(ka/2)+\gamma)$ as the first term in the 
effective range expansion \cite{verhaar}.  This definition
of the scattering length is effectively energy independent once in 
the thresholds regime, $Dk<1$.

Using these wavefunctions to study the basic molecular properties, we 
start 
in the large $a$ limit where the binding energy goes to zero.  At moderate 
$\rho$, the  wavefunction is essentially the same whether the two
particles are a near zero energy scattering state or a loosely bound 
molecule.  Using this fact,  we match log-derivatives 
of the near-zero energy scattering wavefunction in terms of the scattering
length from the short range to the long range
asymptotic bound wavefunction, $\phi_b(\kappa\rho)$.  Using 
the small argument expansions of both wavefunctions allows us to determine 
$\kappa$ in terms of $a$: $\kappa_a=2e^{-\gamma}/a$ and the binding energy 
follows: $-4\hbar^2e^{-2\gamma}/2\mu a^2$ \cite{kanj}. 
Using $\phi_b(\kappa_a\rho)$ will offer us many interesting 
analytic properties of the molecules in the large $a$ limit.  When compared 
to the full multi-channel numerical calculation, these analytic results
provide very good estimates of the molecular properties for large values of $a/D$.

\section{Scattering}
Now we look at the scattering properties of this system as a function of
the polarization angle.  To do this, we solve Eq. (\ref{TISE}) with the Johnson 
Log-derivative propagator \cite{johnson}.  We then extract the T-matrix, 
$T_{if}$, which describes the scattering between channels $i$ and $f$.
The total cross section is $\sigma={1\over k}\sum_{if}|T_{if}|^2$. 
The elastic cross section for $m$ can be written as
$\sigma_m={4\over k}\sin^2(\delta_m)$, where $\delta_m$ is the scattering 
phaseshift for the $m$ partial wave \cite{gu,ajp_2d}.
Sometimes, the most useful quantity is not the scattering cross 
section, rather it is the dimensionless scattering rate $k\sigma$.
Plotting this quantity reveals the system independent or universal
scattering characteristics of the scattering dipolar system, 
as was shown in 3D by Refs. \cite{universal,NJP,roudnev}.
We now present general trends of the 2D dipolar system with a tilted 
polarization axis. 

\begin{figure}
\vspace{3mm}\centerline{\epsfysize=50mm\epsffile{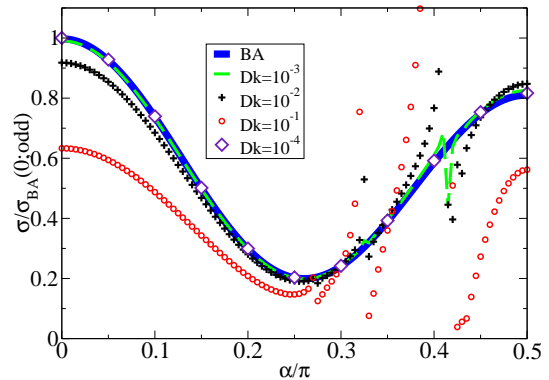}}
\caption{(Color Online) The Born approximation (BA) for fermions shown
as a function of $\alpha/\pi$.  We have plotted the energy independent: 
$\sigma/\sigma_{BA}^{odd}(\alpha=0)$ as a function of $\alpha$ 
for $Dk=10^{-1}$ (red open circle), $10^{-2}$ (blue +), 
$10^{-3}$ (green dash),  $10^{-4}$ (purple open diamond), and the BA is 
shown as a bold blue line.
}\label{BORN}
\end{figure}

In the threshold regime the Born approximation (BA) offers a good estimate of
the scattering for non-zero partial waves \cite{ticknor2d,ajp_2d}.  This 
is  most useful for estimating the cross section for identical fermions. 
In this model the dipoles are spinless
and the way one models fermions or bosons is by imposing the symmetric or anti-symmetric
requirement on the spatial wavefunction.  This leads to the scattering properties of 
fermions being a sum of only the odd partial waves and bosons a sum of only the even.  
For the case of distinguishable particles, one sums up all of the partial waves. 
The BA result for this systems is:
\begin{eqnarray}
&&k\sigma_{BA}^{m\rightarrow m}={4(Dk)^2\over (m^2-{1\over4})^2}\left(1-{3\over2}\sin(\alpha)^2\right)^2 \\\nonumber
&&k\sigma_{BA}^{\pm1\rightarrow \mp1}={4(Dk)^2\over (m^2-{1\over4})^2} \left({3\over4}\sin(\alpha)^2\right)^2 \\\nonumber
&&k\sigma_{BA}^{m\rightarrow m+2}={4(Dk)^2\over\left((m+{1\over2})(m+{3\over2})\right)^2}\left({3\over4}\sin(\alpha)^2\right)^2 \label{BA}
\end{eqnarray}
There are two basic types of collisions given here: diagonal and 
off-diagonal or $m$ changing. For the off-diagonal scattering, there is 
a special case of p-wave collisions where $\pm1$ goes to $\mp1$ and 
has the same functional form 
as the diagonal contribution.  Then for the other off-diagonal terms 
$m\rightarrow m+2$, there is a distinct form.
For identical fermions, the BA can be compared directly to the 
full scattering cross section with out worrying about a short range 
phase shift.  This comparison is made in Fig. \ref{BORN} where the full 
cross section is divided 
by the $\alpha=0$ BA cross section.  Plotting this removes the energy 
dependence of the cross section.  The BA is shown as a thick blue line 
normalized by its $\alpha=0$ value, and the full cross sections 
ares shown for $Dk=10^{-4}$ (violet open diamonds), $10^{-3}$ (dashed green), 
$10^{-2}$ (black $+$), and $0.1$ (red open circles).  Relating
$Dk$ to scattering energy is simply: $E/E_D=(Dk)^2/2$.

In Fig. \ref{BORN} the agreement is good for the full range of $\alpha$, 
especially at small $k$.  But it is worth noticing for $\rho_0/D=0.01$ 
there are two resonances as $\alpha/\pi$ goes to $1/2$.  This are most clearly 
seen in $k=0.1$. For the smaller values of $k$ they are narrow and the 
region where the $\sigma$s deviates from the BA are increasingly small.

\begin{figure}
\vspace{3mm}\centerline{\epsfysize=50mm\epsffile{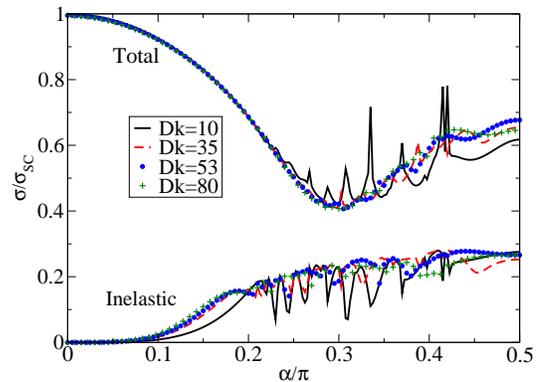}}
\caption{(Color Online) The total and inelastic 
$\sigma/\sigma_{SC}$ as a function of $\alpha$ is
plotted for four different values of $Dk$: 10 (black solid line), 
35 (red dashed), 53 (blue circles), and 80 (green +).
The inelastic data is when $m$ is changed 
in a collisions, and in this case, it makes up 
a significant fraction of the scattering. 
}\label{sc}
\end{figure}

Alternatively, in the high energy regime we can estimate 
the cross section with the Eikonal Approximation \cite{ticknor2d}: 
$\sigma_{SC}={4\over k}\sqrt{\pi Dk}$.  This offers a good estimate of the 
scattering cross section.  We have plotted the scattering cross section over 
$\sigma_{SC}$.  Plotting this quantity removes the energy dependence of the 
scattering.  In Fig. \ref{sc}  we have plotted both the total and inelastic 
($m$ changing) cross sections over $\sigma_{SC}$.
The different energies are  $k$=10 (black solid line), 35 (red dashed), 
53 (blue circles), and 80 (green +).  
This estimate is given for the distinguishable case.  In the case of bosons
or fermions their $\sigma$ will oscillate about $\sigma_{Eik}$
 \cite{ticknor2d,jose}.
Notice that the elastic scattering never turns off even though there is no 
diagonal interaction at $\alpha_c/\pi\sim0.30$.  This shows that the 
scattering is made up from second-order processes; even though there is no 
diagonal interaction, there is a significant diagonal scattering 
contribution because of the off-diagonal channel couplings. 
The shape of this curve is that the total scattering rate dips 
and reduces to about 40$\%$ of it original value then increases up to about 
65$\%$. The inelastic rate starts at zero and quickly climbs until
$\alpha/\pi>0.2$, but after that it only slightly increases.
For the lower end of this regime ($k=10$) scattering, there are still 
noticeable resonances in the scattering when a single partial wave makes up 
a large percentage of the scattering. 

\begin{figure}
\vspace{3mm}\centerline{\epsfysize=90mm\epsffile{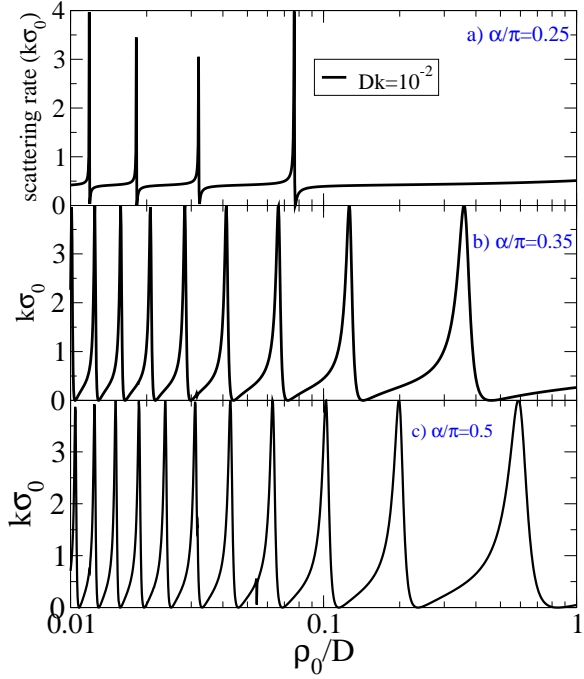}}
\caption{(Color Online) 
The scattering rate, $k\sigma_0$, as a function of $\rho_0/D$ at three 
different angles: a) $\alpha/\pi=0.25$, b) $\alpha/\pi=0.35$, and c) $
\alpha/\pi=0.50$ at $Dk=10^{-2}$ (black). Notice that these resonances for
small angles are narrow and fewer in number than the case where polarization is 
in-plane.  In that case there are many resonances and they are wider in 
$\rho_0/D$.
}\label{angle}
\end{figure}

With these simple estimates of the scattering magnitude in hand, we now move 
to study the impact of the tilted polarization axis on the scattering for 
bosons or distinguishable dipoles, where there is the $m=0$ contribution
relaying information about the short range scattering. For fermions, the
short-range is strongly shielded, and only when one considers specific cases
does the scattering become more involved.  For this reason,
we are more interested in the bosonic scattering and general long-range 
behavior and leave the case specific fermionic scattering for the future.

As a first look at this scattering behavior, we look at the scattering rate
 as a function of $\rho_0/D$ at three different angles: 
a) $\alpha/\pi=0.25$, b) $\alpha/\pi=0.35$, and c) $\alpha/\pi=0.50$.
In Fig. \ref{angle} a) the resonances which occur as $\rho_0/D$ is 
decreased are narrow.  Only a few exist because there is a dipolar barrier 
to the scattering and thus the inter-channel couplings
at short range must be strong enough to support the bound state. 
As the polarization angle is increased, the resonances become wider and more 
frequent.   As the barrier is turned off and ultimately turning into an 
attractive potential, we see more frequent and wide resonances. These 
resonances are much like the long 
range resonances seen in 3D \cite{You,CTPR,NJP,universal,roudnev,kanj2}.

\begin{figure}
\vspace{3mm}\centerline{\epsfysize=60mm\epsffile{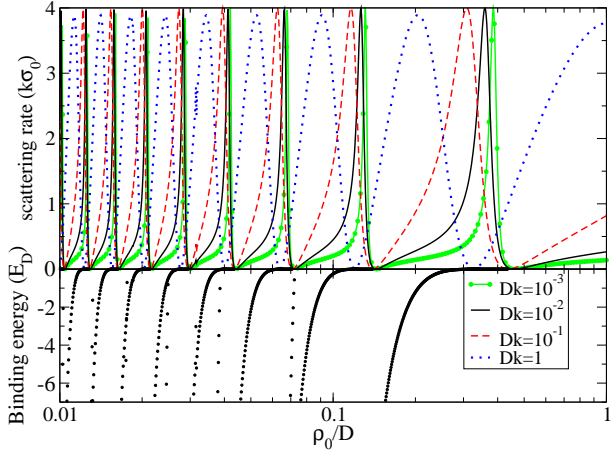}}
\caption{(Color Online) Energy dependence of the 
scattering rate, $k\sigma_0$ and the binding energies are shown 
as a function of $\rho_0/D$ for $\alpha/\pi=0.35$. The curves are for
$Dk=10^{-3}$ (green with circles), $10^{-2}$ (black), 
$10^{-1}$ (dashed red) and $1$ (blue dotted).  The binding energies 
are also shown for the same $\rho_0/D$.  The cross sections and 
binding energy go to zero simultaneously and the strong energy dependence
of the scattering resonance width.
}\label{scattering}
\end{figure}

To study this energy dependence of the scattering further, we
look at a particular angle and vary the energy. 
In Fig. \ref{scattering} we have plotted the scattering rate 
($\alpha/\pi=0.35$) in the upper panel
at four values of $Dk$: $10^{-3}$ (green with 
circles), $10^{-2}$ (black), $10^{-1}$ (dashed red) and $1$ (blue dotted).  
In the lower panel we have plotted the binding energy. 
There are a few points to be made here about the strong energy dependence 
of the scattering. First, the peak of the resonance shifts noticeably as the 
energy is lowered.  Second, the width of the resonance becomes more narrow 
as the energy is decreased. Third, as the the binding energy goes to zero, 
the scattering rate goes to zero; this is in contrast to 3D. 
Fourth, for the $|m|>0$ resonances are very narrow, and they bind tightly as 
$\rho_0/D$ is decreased.  In these plots these resonance are hard to see
because they are so narrow.  They are most easily found by looking at the
binding energy where there are steep lines.

It is worth commenting on the relationship between the scattering length
and the cross section and how these two quantities relate to the binding 
energy.
First, for identical bosons in 3D, the cross section is $\sigma\sim 8\pi a^2$ 
but for $a\gg k$ it saturates at $8\pi/k^2$.  In the large $a$ limit, the 
binding energy goes as $\left(-{\hbar^2\over2\mu a^2}\right)$, and the
maximum of $a$ corresponds to the maximum of $\sigma$. 2D is very different. 
Consider when $a\rightarrow\infty$, and the scattering cross section goes to 
zero.  This is most easily seen as from 
the effective range expansion: 
$\cot(\delta)\propto\ln(ka)\rightarrow\infty$. 
This leads to $\delta\sim0$ and $\sigma\sim0$ when $a$ is very large.
The maximum of the scattering cross section occurs at $\delta=\pi/2$, 
and this happens when $ak=2e^{-\gamma}\sim1.12$. 

\begin{figure}
\vspace{1mm}\centerline{\epsfysize=60mm\epsffile{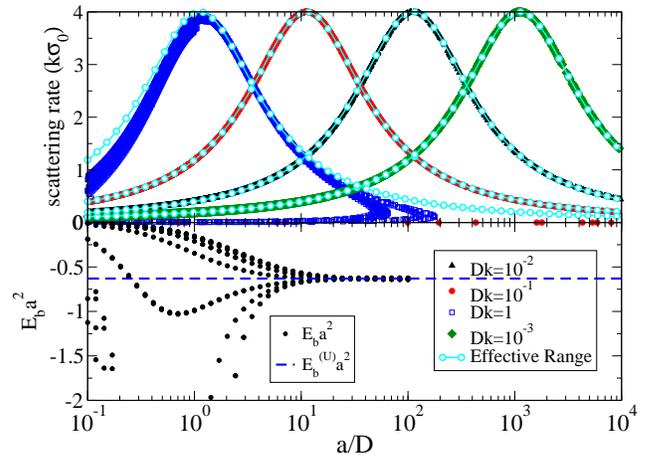}}
\caption{(Color Online)  The scattering rate, $k\sigma_0$, and $a^2E_b$
are shown as a function of $a$ for many values of $\alpha$.
The different energies are:
$Dk=10^{-3}$ (green), $10^{-2}$ (black) $10^{-1}$ (red) and 
$1$ (blue).  Cyan open circles are the effective range at each energy.
The energy dependence of the scattering rate is seen simply
as where the scattering rate peaks when $ka=2e^{-\gamma}\sim1.12$.
The binding energies times $a^2$ are also shown for all resonances,
the blue dashed line is the universal limit, $\sim0.63$.  Only for the 
$Dk=1$ does the effective range not give a good estimate of the scattering 
rate.  }\label{asc_rate}
\end{figure}

A way to clearly demonstrate the energy dependent behavior is shown 
in Fig. \ref{asc_rate}, where we have replotted the scattering rate and 
binding energy as a function of $a/D$, not $\rho_0/D$.
We have replotted all the scattering data
(i.e. many different values of $\alpha$)
as a function of $a/D$ for many different values of $Dk$: 
$10^{-3}$ (black), $10^{-2}$ (red) $10^{-1}$ (green) and 
$1$ (blue).  The binding energies are now plotted as $a^2E_b$, this makes 
it so that when the energies become universal they go to a constant value 
of $\sim$-0.63.  Replotting the data this way, allows us to observe clearly 
several points.  First, the scattering rate is maximum when 
$ka=2e^{-\gamma}\sim1.12$;
this is clear from the four different energies plotted. 
It is also clear that the 2D system has strong energy dependence, and that 
the scattering rate goes to zero in the large $a$ limit.  

From this plot we can understand why there is such strong energy dependence 
to the width of the resonances as a function of $\rho_0/D$.
$a/D$ is energy independent once one is in the threshold regime
and depends only on $\rho_0/D$. 
We know that the scattering rate is zero when $a\rightarrow\infty$ and
that it is maximum when $ka\sim1.12$. Therefore as energy is decreased, 
the maximum and minimum of the scattering rate approach each other in
$a/D$ (Fig. \ref{asc_rate}) or $\rho_0/D$ (Fig \ref{scattering}).

The effective range expansion gives a very good estimate of the 
scattering rate and therefore the phase shift at low $k$.
For $k$=1 (blue squares), we are leaving the threshold regime, and 
the effective range description breaks down. 

Moving to the binding energies, we are see 
$a^2E_b$ (black circles) converges to the universal value (blue 
dashed line) for $a/D>10$ and when it is strongly system dependent.  
We also see that at small $a/D$ the binding 
energies deviate from the universal trend and $a^2E_b$ widely varies. In 
this figure we have only plotted the binding energies which 
were numerically found.  Going beyond $a/D$=100 is both computationally 
challenging and only returns the universal binding energy.

\section{Molecules}

In this system the molecules have widely varying properties depending on 
the polarization angle.  To study them more closely we pick six angles to 
explore: $\alpha/\pi$=0.25, 0.275, 0.3, 0.35, 0.4, and 0.5.
For $\alpha/\pi$ smaller than 0.2, very little variation in the 
scattering, and there are no bound states for the values of 
$\rho_0/D$ we consider.
The first important point is that for a given angle the properties are 
robust.  This is demonstrated by obtaining the molecular energies and 
wavefunctions for the first two $m=0$ resonances for each angle.
Then we determine the values of $\rho_0/D$ at 
each resonance which result in a set of chosen scattering lengths.
We pick 40 different scattering lengths between 0.1 and 
100$D$ that are found at each of the first 2 resonances for each of the 6 
angles.  

This idea of using $a/D$ as the control parameter was used in 3D dipolar 
scattering to study three body recombination of dipoles \cite{3body}.  
That work showed that using  the scattering length to characterize the 
2-body system in the calculations, even outside the $a\gg D$ regime, worked 
well at revealing universal characteristics of three body recombination.

\begin{figure}
\vspace{2mm}
\centerline{\epsfysize=60mm\epsffile{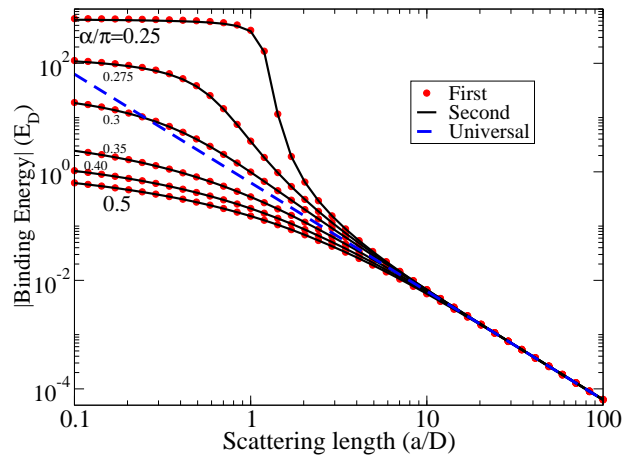}}
\caption{(Color Online) The binding energies for the first and second $m=0$
resonances at $\alpha/\pi$= 0.25, 0.275, 0.3, 0.35, 0.4, and 0.5.  The 
universal 2D binding energy at large $a/D$ is shown as a blue dashed line.
}\label{binding}
\end{figure}

In fig. \ref{binding} we plot the binding energies of the molecules for the 
first (black line) and second (red circle) resonance for all values of 
$\alpha$ considered.  For $\alpha/\pi=0.25$ the molecules are 
most tightly bound at small $a/D$ and at $\alpha/\pi=0.5$ are most loosely 
bound, as expected, although there is roughly 2 orders of magnitude 
difference in the binding energies between these two extreme cases. 
For the tightly bound case the binding energies plateau as $a/D$ is lowered.
This energy corresponds to the size of the hard sphere and 
therefore a minimum size of the molecule.  In this case there is a strong 
dipolar barrier and it is the inter-channel couplings 
that form the attractive short range region where the molecule is found. 
In contrast for $\alpha/\pi=0.5$ there is an attractive dipolar term for 
the  $m=0$ case and the molecules are only mildly multi-channel objects.
This will be explained below.

As $a/D$ is increased, for all of the polarization 
angles, the system becomes more loosely bound, and they are strongly 
system dependent. Only for relatively large $a/D$, say 10, do the binding 
energies truly resemble the universal value. 
We have plotted the universal binding energy as a blue dashed 
line.
\begin{figure}
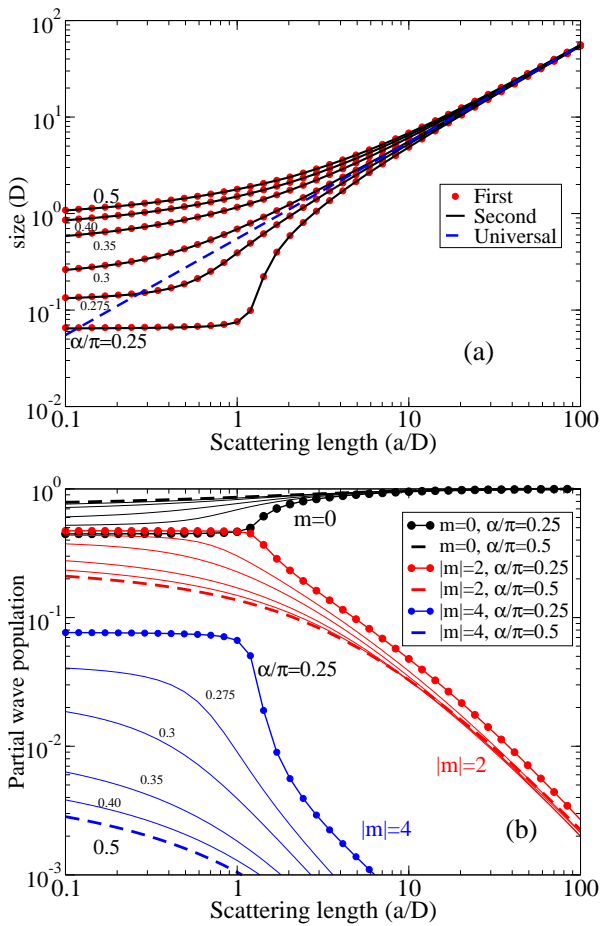

\vspace{2mm}
\centerline{\epsfysize=60mm\epsffile{fig7a.eps}}\vspace{2mm}
\centerline{\epsfysize=60mm\epsffile{fig7b.eps}}
\caption{(Color Online) (a) The size of the molecules are shown as a 
function of $a/D$ for various angles for both the first (black) and second 
(red circles) 
resonance.  The universal form of the size is shown as a dashed blue line. 
(b) The partial wave population is shown (first resonance only), 
for all six angles as a function of $a/D$.
The extreme angles are shown as filled circles ($\alpha/\pi=0.25$) 
and dashed lines ($\alpha/\pi=0.5$) for partial waves $m=0$ (black), 
$|m|=2$ (red) and $|m|=4$ (blue). 
}\label{molprop}
\end{figure}

We now consider the size and shape of the molecules.  First in Fig. 
\ref{molprop} (a) we look at the expectation value of the molecular size:
$\langle \psi|\rho|\psi\rangle$ as a function of scattering length.
All values of $\alpha$ are plotted for both the first (black line) 
and second (red circle) resonance. For $\alpha/\pi=0.25$ and small $a/D$
the molecules are very small, $\langle \rho\rangle\sim0.05D$; 
this is roughly the size of the hard sphere when the first bound state
is captured. In contrast $\alpha/\pi=0.5$ the molecule is about
$\langle\rho\rangle\sim1D$ even for $a/D\sim0.1$.
In the large $a/D$ limit we find that the size of the molecule goes to 
$\pi^2e^{\gamma}a/32 \sim0.55a$ (blue dashed) which was obtained from
$\phi_b(\kappa_a\rho)$. Again we find that the size of the molecules 
depends on the polarization until $a/D>10$.

Now we look at the shape of the molecules shown in  Fig. \ref{molprop} (b).
This is done by considering the partial wave population 
$n_m=\langle \phi_m|\phi_m\rangle$ as a function of scattering length for 
each angle. The extreme angles of $\alpha/\pi=0.25$ and $\alpha/\pi=0.5$ 
are shown as filled circles and dashed lines for $m=0$ (black), $|m|=2$ 
(red) and $|m|=4$ (blue).

For $\alpha/\pi=0.25$ and small $a/D$ the molecules are strongly aligned 
along the polarization axis behind the dipolar barrier. This is why they 
are so small and highly aligned. This is seen by the fact that the largest 
contribution 
is from $|m|=2$ (red circles) and $|m|=4$ (blue circles) is nearly 10$\%$ 
of the partial wave population at small scattering length.  
This is from the fact that the dipoles are behind the dipolar barrier and 
the inter-channel couplings are the origin of the molecule.  
The anisotropy at small $a/D$ is still true for
$\alpha/\pi=0.5$, but the $m$=0 contribution is nearly 80$\%$.
This strong contrast is from because of the 
attractive dipolar interaction for the $m=0$ channel only requires a slight 
amount of  inter-channel coupling to form a bound state.
For the 3D case, the $l=0$ molecule is made up of about 
about 60$\%$ s-wave and the rest is essentially d-wave \cite{3body}.

It is important to notice that at large $a/D$ the partial wave populations 
go to very similar values.  The $m=0$ contribution dominates and all other
contributions become small. 
To better understand the shape of the molecules we plot the radial weighted 
molecular densities.

\begin{figure*}
\vspace{1mm}\centerline{\rotatebox{-90}{\epsfysize=140mm\epsffile{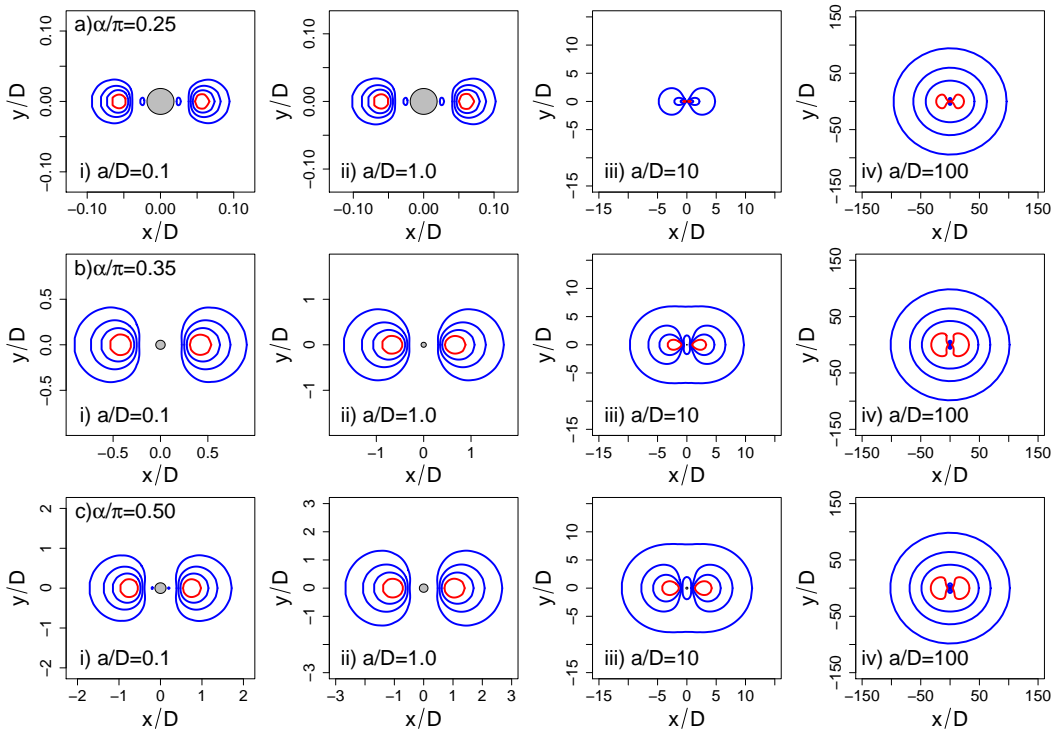}}}
\caption{(Color Online) $\rho |\psi(\vec\rho)|^2$ for:
$\alpha/\pi=$ a) 0.25, b) 0.35, c) 0.5 for $a/D$ (i) 0.1, (ii) 1, (iii) 
10 and (iv) 100.
The contours indicated are drawn ever 20$\%$ of maximum value, with the 
largest 80$\%$ contour being drawn as red.  The solid grey circle in the 
middle is the hard sphere short range interaction.  The scale changes 
for every plot on the left  (i, ii), in contrast the scale is the same for 
the plots on the right (iii, iv) when the system is in the large $a/D$ 
regime. }\label{psi1}
\end{figure*}

In Fig. \ref{psi1} we have plotted the radial weighted molecular densities: 
$\rho |\psi(\vec\rho)|^2=|\sum_m u_m(\rho)e^{im\phi}|^2$
for $\alpha/\pi=$ a) 0.25, b) 0.35, c) 0.5 (top to bottom) for 
$a/D$ of (i) 0.1, (ii) 1, (iii)  10, and (iv) 100 (left to right).
These densities are generated from the second resonance.  The
first resonance wavefunctions look the same expect there inner hard core 
is larger and over takes the inner oscillations.  The scale changes
for the plots on the left. 

This plot clearly shows the change in both shape and size of the molecules
as both $a/D$ and $\alpha/\pi$ are changed.  Now starting at 
$a/D=0.1$ and $\alpha/\pi$=0.25 (ai) we see that the molecule is 
very small and highly anisotropic.  The density is along the 
polarization axis, $x$, and very tightly bound against the hard core. In 
fact its spatial
extend does not really extend beyond the hard core in the $y$ direction. 

Then as the angle is increased, for a fixed $a/D$ the size of the molecule 
and anisotropy is softened. Observe the change is scale in both (bi) and 
(ci).  In (ci), the molecule is larger ($\sim1D$), and still 
aligned along the polarization axis. 
Additionally see the contrast in (ai) and (ci) between the extend of the 
density over the width in $y$ of the hard core.

Now for (a), (b), and (c) consider increasing $a/D$. The size of
the molecules gets larger
and more isotropic.  In (iv) for all angles the system is isotropic, 
except for small region near the hard core, but the bulk of the 
radial weighted density is isotropically distributed at large $\rho$.
This is why the size and shape of the molecules are universal in the large 
$a/D$ limit. 

\section{Conclusions}
In this paper we have studied the scattering properties of the a pure 2D
dipolar system when the polarization can tilt into the plane of motion. 
We have shown how the tilt angle impacts the scattering in both the 
threshold and semi-classical regimes.
We then studied the character of the scattering resonances generated by
altering $\rho_0/D$ or electric field as a function of
the polarization angle.  We found that at small angles the systems
gained bound states which produced narrow resonances.  This is because of the
dipolar barrier.  We also found when the polarization is entirely in the
plane of motion, the resonances are frequent and wide because of the partially 
attractive potential.

We studied the molecular system generated the tilt of the polarization.
We showed that at large $a/D$ the system have a universal shape independent
of polarization angle, but at small $a/D$ we have found that the molecules
have a wide range of properties which strongly depend on polarization 
angle. Future work on this topic will be to more fully consider a 
realistic system, to include the effect of confinement, fermionic dipoles
and layer systems.

\begin{acknowledgments}
The author greatfully acknowledges support from the
Advanced Simulation and Computing Program (ASC) through the 
N. C. Metropolis Fellowship and LANL which is operated by LANS, LLC for the 
NNSA of the U.S. DOE under Contract No. DE-AC52-06NA25396.
\end{acknowledgments}

\end{document}